\documentclass[pdflatex,sn-nature]{sn-jnl}

\usepackage{graphicx}%
\usepackage{multirow}%
\usepackage{amsmath,amssymb,amsfonts}%
\usepackage{amsthm}%
\usepackage{mathrsfs}%
\usepackage[title]{appendix}%
\usepackage{xcolor}%
\usepackage{textcomp}%
\usepackage{manyfoot}%
\usepackage{booktabs}%
\usepackage{algorithm}%
\usepackage{algorithmicx}%
\usepackage{algpseudocode}%
\usepackage{listings}%
\usepackage[version=4]{mhchem}
\usepackage[normalem]{ulem}

\usepackage{multibib}
\newcites{SI}{Supplementary References}

\theoremstyle{thmstyleone}%
\theoremstyle{thmstyletwo}%

\theoremstyle{thmstylethree}%

\usepackage{url}

\AtBeginDocument{
  \renewcommand{\url}[1]{}
}

\newcommand{\name}{TMCgen}

\begin{document}

\title[Article Title]{Manifold Diffusion for Structure Generation of Transition Metal Complexes}


\author[1,2]{\fnm{Luca} \sur{Schaufelberger}}
\author*[1,2]{\fnm{Kjell} \sur{Jorner}}\email{kjell.jorner@chem.ethz.ch}

\affil[1]{Institute of Chemical and Bioengineering, Department of Chemistry and Applied Biosciences, ETH Zurich}

\affil[2]{NCCR Catalysis, Switzerland}


\abstract{

Transition metal complexes are central to catalysis, drug design, and materials science, with relevant properties strongly sensitive to their three-dimensional geometry. However, the electronic diversity and unconventional bonding environments of transition metal complexes pose a major challenge for accurate structure generation.
In this work, we introduce \name{}, a manifold diffusion machine learning model that efficiently and accurately generates geometries of transition metal complexes.
By formulating the diffusion process over the metal-ligand coordination angles, combined with torsional and rotational diffusion of the ligands, \name{} focuses on the key geometric degrees of freedom of transition metal complexes.
\name{} shows strong performance in generating accurate coordination environments on a diverse set of experimentally derived bioinorganic and organometallic complexes while requiring only few inference steps, enabling efficient generation.
Our results demonstrate the potential of manifold-based generative modeling for data-efficient geometry generation, paving the way for property-conditioned design of transition metal complexes.}

\keywords{Generative machine learning, transition metal complexes, manifold diffusion, coordination environments}



\maketitle

\section{Introduction}

\begin{figure}[t!]
    \centering
    \includegraphics[width=\textwidth]{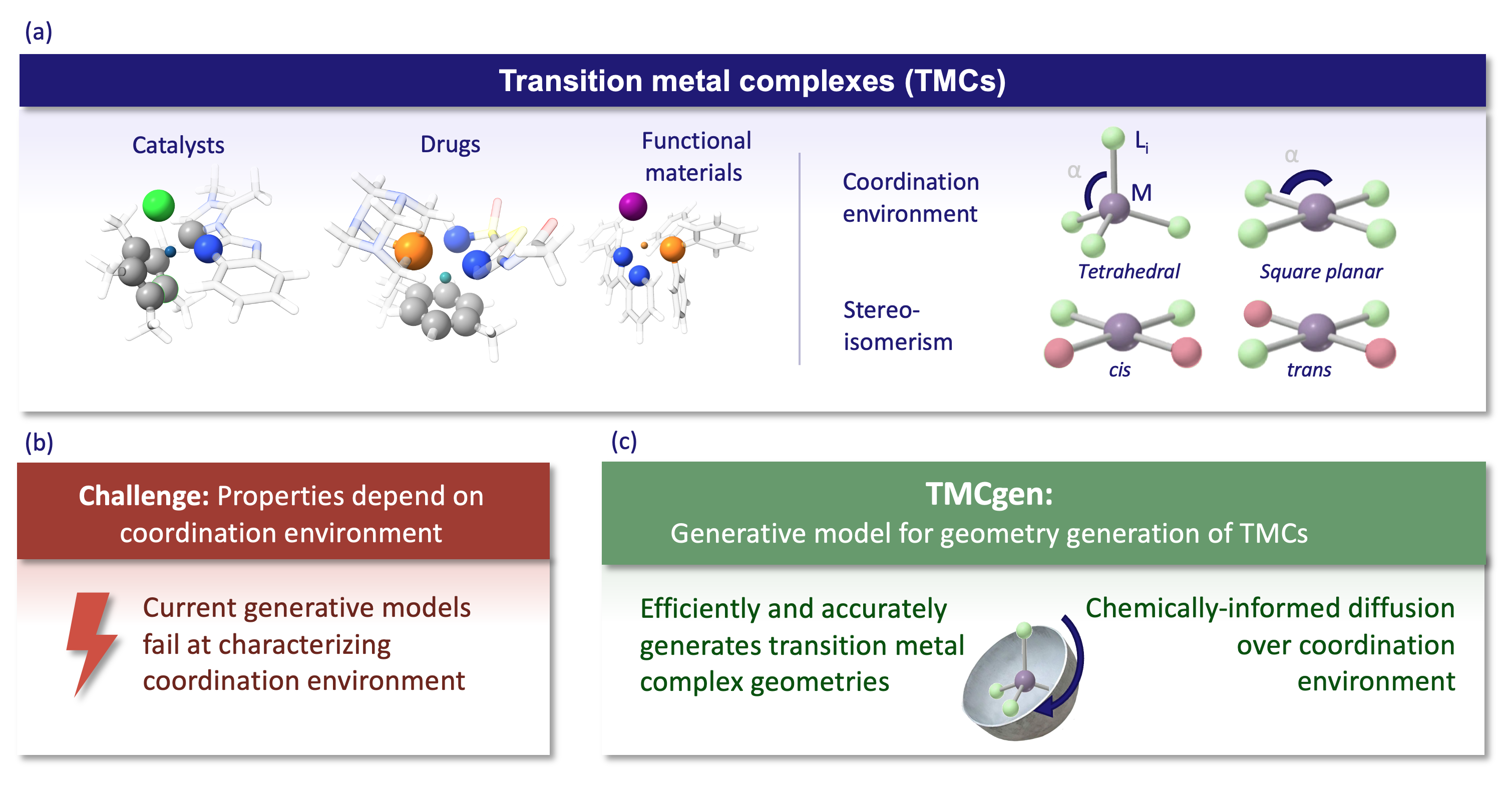}
    \caption{Overview of this work. (a) Left: Examples of transition metal complexes used as catalysts, drugs and functional materials, with the central metal atom and the ligating atoms highlighted. Right: Simplified depiction of two exemplary coordination environments, tetrahedral and square planar (b) Challenges addressed by (c) the manifold diffusion model \name{} introduced in this work.}
    \label{fig:overview}
\end{figure}

Transition metal complexes (TMCs) play a crucial role in catalysis \cite{hartwig_organotransition_2009}, drug design \cite{ ndagi_metal_2017} and materials science \cite{kalyanasundaram_applications_1998} due to their unique structural diversity and electronic properties. 
Monometallic TMCs consist of a central transition metal atom bonded to surrounding molecules or ions, called ligands (see Fig. \ref{fig:overview}a). 
In catalysis, transition metal complexes are crucial to achieving important transformations such as C--H activation \cite{docherty_transition-metal-catalyzed_2023}, asymmetric hydrogenation \cite{wen_asymmetric_2021}, and cross-coupling reactions \cite{jana_advances_2011}, many of which underpin sustainable synthesis and green energy storage technologies \cite{che_lah_late_2021}. 
Transition metal complexes have also found broad application as drugs, where they have demonstrated inhibitory effects on various proteins \cite{bindoli_thioredoxin_2009, shagufta_transition_2020, dorr_metal_2014}. 
The tunable electronic structure of transition metal complexes enables their use in functional materials, for example in spintronics \cite{coronado_molecular_2016} and molecular electronics \cite{amin_review_2023}. 

The unique properties of TMCs are determined by the three-dimensional arrangement of ligands surrounding the metal center. Specifically, both the metal coordination environment, \textit{i.e.}, the angular distribution of ligands around the metal center (\textit{e.g.}, tetrahedral or square planar, see Fig. \ref{fig:overview}a), and the relative positioning of ligands (\textit{e.g.}, \textit{cis} or \textit{trans}) strongly affect their electronic structure. 
For example, for \ce{Ni^{2+}} ($d^8$) complexes, strong-field ligands favor square planar geometries (\textit{e.g.}, \ce{[Ni(CN)4]^2-}), which are low-spin and diamagnetic, whereas weak-field ligands favor tetrahedral geometries (\textit{e.g.}, \ce{[NiCl4]^2-}), which are high-spin with two unpaired electrons and therefore paramagnetic \cite{hartwig_organotransition_2009}.

To achieve a large-scale computational design of transition metal complexes \cite{strandgaard_deep_2025}, accurate generation of their three-dimensional geometries is essential.
However, these geometries are heavily modulated by steric and electronic ligand effects that cause them to often deviate significantly from idealized coordination environments \cite{ketkaew_octadist_2021}. For instance, bulky ligands tend to adopt positions that minimize steric hindrance, while multidentate ligands  --- those with multiple metal-binding sites --- can impose rigid geometries due to the constrained bite angle.
Due to these complexities, generating realistic coordination geometries for TMCs remains highly challenging. 

Traditional approaches to 3D-geometry generation include metadynamics methods \cite{pracht_crestprogram_2024}, which however have limited usability in large-scale discovery campaigns due to their computational expense. Knowledge-based cheminformatics methods such as RDKit's ETKDG algorithm \cite{riniker_better_2015} have been primarily designed for generating conformers of organic molecules and exhibit limited applicability for TMCs, as they do not incorporate experimentally derived coordination-angle preferences for TMCs analogously to how torsional preferences are incorporated for organic molecules.
Instead, they use default coordination environments and position ligands randomly on the available coordination sites if this has not been defined a priori in the input. 
Among approaches that produce transition metal complex geometries 
\cite{sobez_molassembler_2020, taylor_architector_2023, chernyshov_mace_2024, lee_metallogen_2025, toney_graph_2025, clarke_dart_2024}, molSimplify \cite{ioannidis_molSimplify_2016}, which combines cheminformatics methods with classical force fields,  has been widely used for automatic screening of TMCs.
However, it requires prior input of the coordination environment, and either places ligands randomly at coordination sites or based on explicit input, limiting its effectiveness for targeted design --- particularly since TMCs often possess numerous stereoisomers. 

Generative models such as those based on diffusion have recently surpassed the performance of cheminformatics tools for 3D-geometry generation in terms of precision and coverage of the conformational space  \cite{jing_torsional_2022,wang_swallowing_2024} , while being sufficiently fast to be used in large-scale discovery campaigns. 
They further allow for conditional geometry generation based on desired target properties \cite{ho_classifier-free_2022,weiss_guided_2023}, enabling inverse design workflows wherein molecular geometries are generated to meet specific functional or electronic criteria. Generative models have also been used to design new ligands or whole transition metal complexes \cite{jin_liganddiff_2024, cornet_om-diff_2024, strandgaard_deep_2025}.  

In chemistry and biology, diffusion models fall into two categories: Euclidean diffusion models, which operate directly on 3D atomic positions, and manifold diffusion models, which operate on a lower-dimensional manifold, \textit{i.e.}, a chemically-relevant coordinate space such as torsion angles
\cite{jing_torsional_2022, corso_diffdock_2023, yim_diffusion_2024, xu_geodiff_2022, schaufelberger_generating_2026}.
Euclidean diffusion models provide flexibility but rely on large and diverse training datasets, which are not available in many target domains such as transition metal complexes. For instance, while the GEOM dataset for small organic molecules \cite{axelrod_geom_2022} contains 37 million conformers, the experimentally derived TMC dataset \cite{kneiding_deep_2023} used in this work comprises only 61 thousand structures, \textit{i.e.}, nearly three orders of magnitude fewer.
On the other hand, manifold generative models incorporate domain-specific information by limiting diffusion to the most chemically meaningful degrees of freedom, leading to data-efficient learning and providing guarantees for chemical validity, \textit{e.g.} 
reasonable bond lengths \cite{jing_torsional_2022, corso_diffdock_2023, yim_diffusion_2024,elhag_manifold_2024, schaufelberger_generating_2026}. For example, diffusion over torsion angles exhibits strong performance by reducing the degrees of freedom, which are the most important for most small organic molecules. Recently, the PuckerFlow model introduced by us has demonstrated the strength of manifold generative models in the accurate generation of cyclic conformers \cite{schaufelberger_generating_2026}.
Manifold generative models have also been shown to allow for sampling with few inference steps, enabling compute-efficient generation \cite{jing_torsional_2022}.

Despite these advances, current manifold diffusion models are limited to organic molecules, as they address insufficient degrees of freedom to generate the coordination environments of transition metal complexes. For example, the approach used in torsional diffusion keeps all bond lengths and angles fixed, including the coordination bonds and angles of ligands around a metal center.
Therefore, new methodologies are required that can model the coordination environments of transition metal complexes effectively.

Here, we introduce \name{}, a manifold diffusion model that achieves strong performance in generating three-dimensional geometries of transition metal complexes.
We demonstrate the performance of  \name{} on a highly diverse set of transition metal complexes extracted from experimental crystal structures, on which our model generates geometries with accurate quantum-mechanical properties. 
By constraining the diffusion process over the relevant manifold, \textit{i.e.}, the angular distribution of ligands around the metal center combined with ligand rotations and internal torsion angles, \name{} shows strong performance with few inference steps.
 \name{} lays the foundation for conditional geometry generation of metal complexes in inverse design settings, enabling the targeted exploration of structures with desired characteristics for drug discovery and sustainable catalyst development.

\section{Methods}

\begin{figure}[t!]
    \centering
    \includegraphics[width=\textwidth]{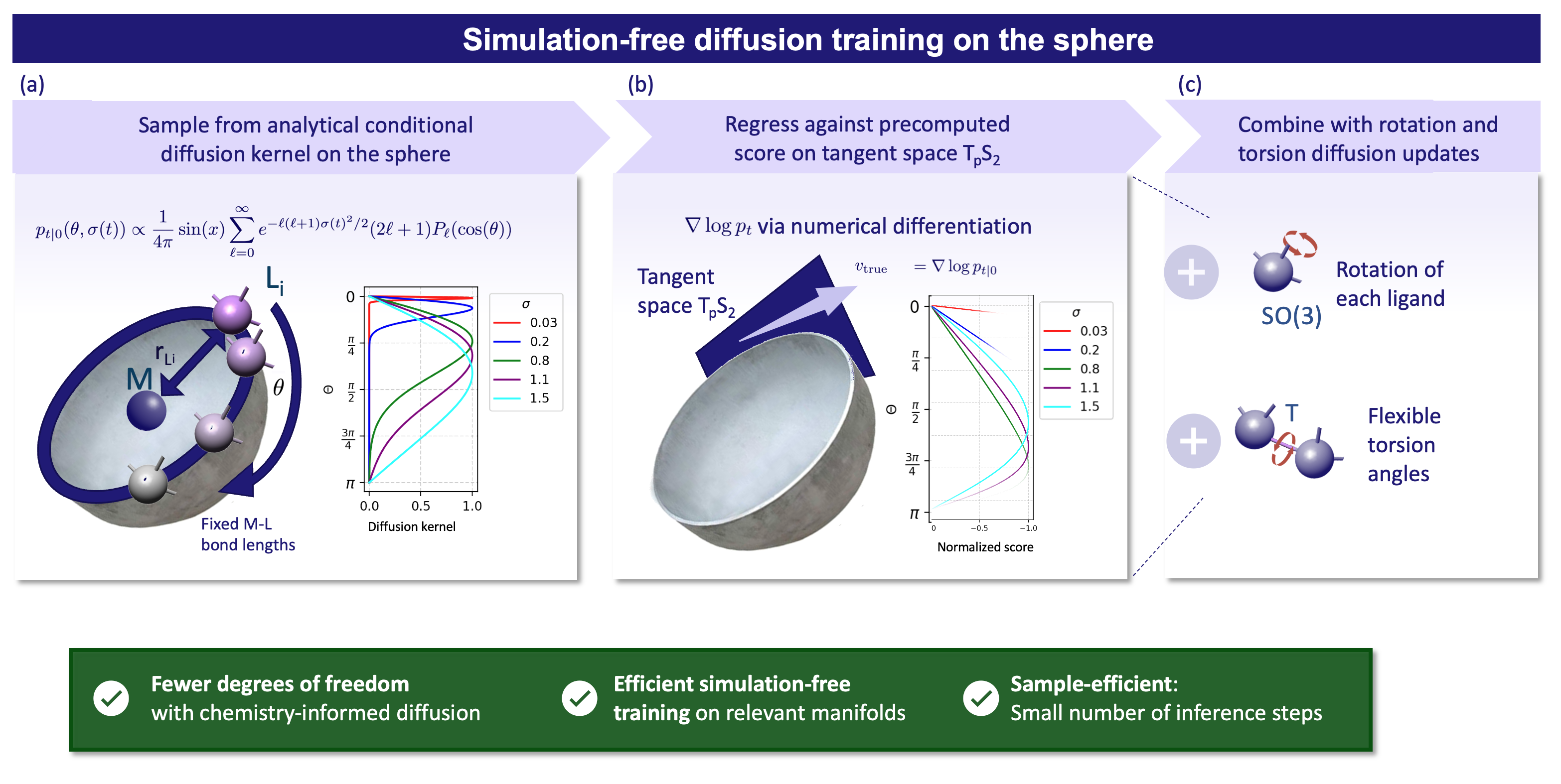}
    \caption{Simulation-free manifold diffusion on the sphere. (a) Sampling from the analytical conditional diffusion kernel on $S^2$ using a closed-form heat kernel expansion while keeping metal-ligand bond lengths fixed. (b) Training diffusion model via regression to the score, precomputed by numerical differentiation. The model output is projected onto the tangent space. (c) Coupling with ligand rotations and torsional updates to model full transition metal complex degrees of freedom.}
    \label{fig:method_sphere}
\end{figure}

 \name{} formulates the geometry generation task for transition metal complexes via manifold diffusion, \textit{i.e.}, it operates in an internal coordinate system reducing the number of degrees of freedom.
The most relevant degree of freedom of TMCs is the coordination geometry around the central metal atom, which is characterized by the relative angles between the ligands and their distance to the central metal atom (see Fig. \ref{fig:overview}a).
Therefore, we formulate the diffusion process on the angular distribution of the ligands around the metal center, corresponding to the surfaces of a set of spheres centered on the metal atom and with radii equal to the corresponding metal-ligand bond lengths for each ligand. 
We introduce a diffusion on the sphere based on the explicit spherical diffusion kernel and combine it with the well-established diffusion over rotational and torsional degrees of freedom of each ligand (see Fig. \ref{fig:method_sphere}). 
Unlike previous manifold diffusion models on the sphere for other application domains \cite{de_bortoli_riemannian_2022}, our approach is simulation-free, \textit{i.e.}, it avoids solving the diffusion stochastic differential equation during training (see Secs. \ref{sec:methods_sim_free} and \ref{sec:methods_tmc_sphere_diffusion}).  
Specifically, we apply updates to the diffusion over the coordination environment, the rotations and the torsions for each ligand separately, and design our architecture to handle an arbitrary number of ligands and torsion angles (see Fig. \ref{fig:method_coupling_DOFs} and Sec. \ref{sec:methods_coupling_with_DOFs}).

\subsection{Simulation-free Manifold Diffusion}\label{sec:methods_sim_free}
Diffusion generative models consider a forward noising process that corrupts the training data distribution  \( p(x_0) \) via the stochastic differential equation (SDE): 
\begin{equation} 
dx = \sqrt{d\sigma^2(t)/dt} \, dw 
\label{eq:methods_forward_diffusion}
\end{equation}
where \( w \) is the corresponding Brownian motion and $\sigma(t)$ a predefined noise schedule. The model then learns the \textit{score} $  \nabla \log p_t$ of the noisy data distribution, with which it is possible to generate new samples via the reverse SDE starting from a simple noise distribution:
 \begin{equation} dx = [ - g^2(t) \nabla_x \log p_t(x)] dt + g(t) d\bar{w}. \end{equation}

where $g(t) = \sqrt{d\sigma^2(t)/dt}$ and $\bar{w}$ is a reverse-time Brownian motion.

To train the score model, a procedure is required to generate samples $x_t$ with different amounts of noise added and compute their conditional score $  \nabla \log p_{t \mid 0}(x_t \mid x_0)$, which serves as the regression target.
This can either be performed by simulation, \textit{i.e.}, explicitly solving the forward SDE (Eq. \ref{eq:methods_forward_diffusion}) during each training step, or analytically by adding all the noise corresponding to a certain time step at once. More formally, the latter simulation-free approach requires to sample directly from the conditional diffusion kernel $p_{t \mid 0}(x_t  \mid x_0)$ and to compute its conditional score $  \nabla \log p_{t \mid 0}(x_t \mid x_0)$. 
%
In Euclidean diffusion, the diffusion kernel is a Gaussian around the data point $x_0$, making simulation-free training straightforward.

However, formulating diffusion on non-Euclidean spaces relevant in many chemical or biological applications requires a generalization to Riemannian manifolds, which intuitively correspond to smoothly curved spaces equipped with a local notion of distance, allowing to define \textit{e.g.}, angles and curvature \cite{lee_introduction_2018}. 
Examples of Riemannian manifolds include the \textit{n}-sphere, ellipsoids, paraboloids and certain Lie groups such as SO(n). 
On many manifolds, the diffusion kernel $p_{t \mid 0}$ is not easily accessible, and $  \nabla \log p_{t \mid 0}$ needs to be calculated on the tangent space, which is a linear approximation at a given point on the manifold.
In chemistry and biology, simulation-free training has been developed for the torus $\mathbb{T}$ (parametrizing molecular torsion angles), and the group SO(3) (parametrizing rotations), where the diffusion kernel is given by the wrapped normal and IGSO(3) distributions, respectively \cite{jing_torsional_2022, leach_denoising_2022, yim_diffusion_2024}. 
For example, the molecular docking model DiffDock \cite{corso_diffdock_2023} treats the diffusion process over ligand translations, rotations and torsion angles, pre-computing the diffusion kernels and their corresponding scores for efficient training. 
Building on these ideas, we extend manifold diffusion to transition metal complexes, which involve structurally diverse coordination geometries not addressed by previous manifold diffusion models.
\subsection{Manifold Diffusion for Transition Metal Complexes}\label{sec:methods_tmc_sphere_diffusion}
\begin{figure}[t!]
    \centering
    \includegraphics[width=1\textwidth]{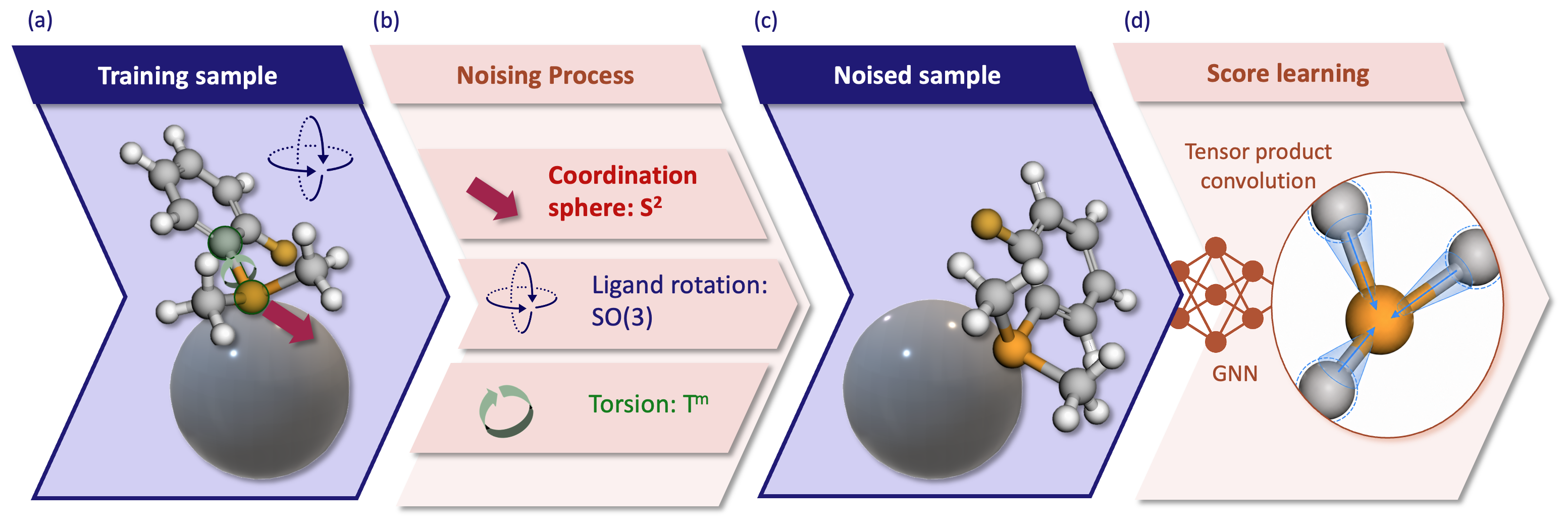}
   \caption{Coupled manifold diffusion for transition metal complexes. (a) Ground-truth structure, illustrating the noising on coordination sphere (red), ligand rotations (blue), and torsions (green). For ease of visualization, only a single ligand is shown. (b) Independent noising on the product manifold $S^2 \times \mathrm{SO}(3) \times \mathbb{T}^m$ producing the (c) noised sample. (d) Score learning with an equivariant GNN, predicting tangent-space updates for each degree of freedom via tensor product convolutions.}
    \label{fig:method_coupling_DOFs}
\end{figure}

In this work, we formulate the diffusion process over the angular distribution of the ligands around the metal center corresponding to the sphere $S^2$, and pair it with the established manifold diffusion over rotations and torsion angles (see Fig. \ref{fig:method_coupling_DOFs}).
Diffusion models on the sphere have been formulated by De Bortoli et al. \cite{de_bortoli_riemannian_2022} in other domains in terms of projecting a geodesic random walk in $\mathbb{R}^3$ space onto the sphere.
However, the forward diffusion process (Eq. \ref{eq:methods_forward_diffusion}) was explicitly solved by simulation during training, which is impractical for large-scale model training. 

We instead base our training routine on the explicit conditional diffusion kernel $p_{t \mid 0}(\theta, \sigma(t))$, which can be obtained by rescaling the heat kernel $G(\theta, \sigma(t))$ that solves the heat equation on the sphere (see Fig. \ref{fig:method_sphere}a). The heat kernel is given by the uniformly and absolutely convergent power series \cite{zhao_exact_2018}:

\begin{equation}
\label{eq:methods:heat_kernel}
G(\theta, \sigma(t)) =
\frac{1}{4\pi} \sum_{\ell=0}^{\infty} e^{-\ell(\ell+1)\sigma(t)^2 / 2 } (2\ell + 1) P_{\ell}(\cos(\theta))
\end{equation}
where $\theta$ is the (polar) angle between $x_t$ and $x_0$, \textit{i.e.}, $\theta = \arccos\left( \frac{\langle x_t, x_0 \rangle}{\|x_t\| \, \|x_0\|} \right)$ and $P_{\ell}$ is the $\ell$-th Legendre polynomial,  which can be computed with \textit{Bonnet}’s recursion formula.
For practical implementation, we cut the infinite series upon numerical convergence, and use efficient parallelization, leveraging that only the exponential term depends on the noise schedule $\sigma(t)$.  

By symmetry, the diffusion kernel depends only on the polar angle $\theta$ (see Fig. \ref{fig:method_sphere}a), while the azimuthal angle $\phi$ relative to $x_0$ can be sampled uniformly from $[0, 2\pi]$.
While the diffusion modeling framework presented here can be phrased for any $n$-sphere with $n \geq 2$ that might be relevant in specific application domains, such as for the conformational ensembles of molecular ring systems, we focus on the $2$-sphere in the main text. The  generalization to $n > 2$ is presented in SI \ref{si:nsphere}, where the Legendre polynomials are replaced by the more general Gegenbauer polynomials \cite{zhao_exact_2018}.
To sample from the conditional diffusion kernel $p_{t \mid 0}(\theta, \sigma(t))$, $G$ is weighted by the surface area element of the uniform distribution on the sphere:
\begin{equation}
    p_{t \mid 0}(\theta, \sigma(t)) \propto \sin(\theta)     G(\theta, \sigma(t))
    \label{eq:methods:cond_prob}
\end{equation}

During training, we regress the model output against the true score, as visualized in Fig. \ref{fig:method_sphere}b. This corresponds to a SE(3)-equivariant vector on the tangent space of $S^{2}$, i.e., $T_{\mathbf{p}}S^{2}$, given by:
\[
T_{\mathbf{p}}S^{2} = \{ \xi \in \mathbb{R}^{3} : \langle x_t, \xi \rangle = 0 \}
\]
where \( \langle \cdot, \cdot \rangle \) denotes the Euclidean inner product and \( x_t \in S^{2} \). We precompute the magnitude of the true score via numerical differentiation of the logarithm of Eq. \ref{eq:methods:cond_prob}, with the direction defined by the projection of $ x_0 - x_t$ onto $T_{\mathbf{p}}S^{2}$.  During training, we project the model output, an irreducible representation of SO(3) \cite{geiger_e3nn_2022}, onto the tangent space $T_{\mathbf{p}}S^{2}$ and regress against the true score vector.

At inference, we sample from the prior $p(x_1)$, which due to compactness of the manifold is not a Gaussian, but the uniform distribution. We perform a geodesic random walk by performing an Euler-Maruyama step with noise $z_t$ on the tangent space and projecting the resulting tangent vector \( \xi \) back onto the sphere through the exponential map:
\begin{equation}
\exp_{\mathbf{x}}(\xi) = \cos(\lVert \xi \rVert_2)\mathbf{x} + \sin(\lVert \xi \rVert_2)\frac{\xi}{\lVert \xi \rVert_2}
\end{equation}
which intuitively corresponds to a rotation by the magnitude of \( \xi \) along the geodesic. %

\subsubsection{Coupling Sphere Diffusion to Ligand Degrees of Freedom}\label{sec:methods_coupling_with_DOFs}
We combine the diffusion over the spherical coordination environment derived above with manifold diffusion over the ligand rotation and torsion angles (see Fig. \ref{fig:method_sphere}c).
Especially, we separately diffuse over the rotation of each ligand and update the torsion angles of each ligand (see Fig. \ref{fig:method_coupling_DOFs}a), thus ensuring that the relevant degrees of freedom of transition metal complexes are covered.
We adapt the rotational and torsional diffusion methods from \cite{jing_torsional_2022, yim_se3_2023}  to better disentangle the degrees of freedom in transition metal complexes, as shown in Fig. \ref{fig:method_coupling_DOFs}b.
Instead of rotating ligands around their center of mass, we rotate them around the ligating atom, while keeping the metal center stationary. After torsional updates, ligands are realigned to keep the ligating atoms fixed, preserving metal-ligand bond lengths. 
To avoid high variance in the true score due to the metal center's spherical symmetry, one ligand is not rotated, reducing training variance without introducing bias or data leakage.
While our adapted diffusion process ensures feasible metal-ligand bond lengths, it does not guarantee the correct orientation of multidentate ligands, which coordinate to the metal via multiple binding atoms. Therefore, we adjust the rotation of multidentate ligands after the last diffusion step based on the target bond lengths. 

Like previous manifold diffusion models, \name{} requires a reference structure for the \textit{local environment} of each ligand, \textit{i.e.}, intra-ligand bond lengths, bond angles and
cycle conformations. The conformer matching procedure of Jing et al. \cite{jing_torsional_2022} is not directly applicable to transition metal complexes as it treats the metal coordination environment rigidly. Therefore, we train on the ground truth local structure environments, while at inference using the local environments extracted from RDKit's ETKDG algorithm of the ligands in their metal binding configuration, as is common  across other manifold diffusion models such as torsional diffusion \cite{jing_torsional_2022} or DiffDock \cite{corso_diffdock_2023}. 
Alternatively to the local environments sampled from RDKit's ETKDG algorithm, users may separately supply ligand structures as reference local environments. This could, \textit{e.g.}, be structures from the tmQMg-L dataset \cite{kneiding_directional_2024}, or incorporate recent generative modeling of cyclic local environment with PuckerFlow \cite{schaufelberger_generating_2026}. Our model can be used for sampling either by initializing from RDKit geometries (presented in Sec. \ref{sec:results}) or with fully random initialization, \textit{e.g.}, including random rotations of all ligands (see Sec. \ref{si:res_random_init}).  In the former option, \name{} starts with the default RDKit initialization of the coordination geometry and random placements of ligands, and improves on this initialization acting as a structure refinement model.

\subsection{Computational Details}
\subsubsection{Model architecture} 
In \name{}, we employ an equivariant neural network based on \textit{e3nn} \cite{geiger_e3nn_2022} for representation learning, and adapt the model output layers to the problem of generating transition metal complexes.
Specifically, we adapt the output layers to account for the varying dimensionality of the output depending on the number of ligands $N$ and torsion angles $m$, enabling the architecture to scale naturally to any number of ligands.
Our score model outputs $2N$ SE(3)-equivariant vectors in the tangent spaces of translation and rotation (Euler angles), similar to \cite{somnath_dockgame_2023}, and extends it to model the variable number of torsion angles of the ligands.
For each ligand, rotational and translational updates are computed via a tensor product convolution between the learned atomic features and the ligating atom (see Fig. \ref{fig:method_coupling_DOFs}d). 
This allows the model to naturally handle a varying number of ligands across training and inference. 
The score model consists of a four-layer E3NN with 48 scalar and 10 vector channels using spherical harmonics up to $\ell_{\max}=2$. Pairwise distances and diffusion times are encoded through sinusoidal embeddings and processed with tensor-product convolutions over radius graphs with a 5~\AA{} cutoff (see SI \ref{si:hyperparams} for full hyperparameters, and publicly available codebase).

\subsubsection{Data}
We trained our model on the tmQMg dataset \cite{kneiding_deep_2023}, which consists of TMCs derived from experimental data with a large diversity of coordination environments, thus constituting a challenging real-world learning target. The tmQMg dataset includes complexes based on a large variety of organic ligands and all thirty elements from the 3d, 4d, and 5d series. 
As the tmQMg dataset contains only a single geometry per complex, the learning task is different compared to conformer sampling, where multiple geometries are available during training. Generating the coordination environment is also strictly not conformer generation as it can produce different stereoisomers (\textit{e.g.}, \textit{cis}, \textit{trans}) of the same complex.
We preprocess the tmQMg dataset based on the connectivity information provided in the dataset to identify ligands and bonding patterns. Random splitting results in 58,000 train complexes, and 1,400 validation and test set complexes each; with preprocessed data available at the Zenodo repository (see Data Availability).

For benchmarking, two generative models (ConfGF \cite{shi_learning_2021}, GeoDiff \cite{xu_geodiff_2022}) were trained from scratch on the tmQMg training set with the originally reported hyperparameters. RDKit geometries were created using the SMILES extracted with $\text{xyz2mol}_\text{tm}$ \cite{rasmussen_smiles_2025} and the ETKDG algorithm \cite{riniker_better_2015}. For each method, we generated a single molecular geometry per test complex and report the error of geometric and electronic properties relative to those of the reference structure. For the electronic properties, we used GFN2-xTB \cite{ bannwarth_gfn2-xtbaccurate_2019} with SCF convergence of $10^{-6}$ Eh, singlet spin states and an electronic temperature of 300 K. 

\section{Results}\label{sec:results}
\begin{table}[t]
\centering

\small
\setlength{\tabcolsep}{4pt}
\begin{tabular}{lcccccccc}
\toprule
& \multicolumn{2}{c}{RMSE$_{\mathrm{ang.}}$ (rad)} & \multicolumn{4}{c}{Frac. RMSE$_{\mathrm{ang.}}$ $<\tau$} & \multicolumn{2}{c}{RMSD$_{\mathrm{pos.}}$ (\AA)} \\
\cmidrule(lr){2-3} \cmidrule(lr){4-7} \cmidrule(lr){8-9}
Method & Mean(↓) & Med.(↓) & $<0.5$(↑) & $<0.4$(↑) & $<0.3$(↑) & $<0.2$(↑) & Mean(↓) & Med.(↓) \\
\midrule
\textbf{\name{} (ours)} & \textbf{0.43} & \textbf{0.41} & \textbf{0.60} & \textbf{0.49} & \textbf{0.41} & \textbf{0.31} & \underline{2.31} & 2.24 \\
RDKit & 0.63 & 0.66 & 0.27 & 0.17 & 0.10 & 0.04 & 2.37 & 2.28 \\
GeoDiff & \underline{0.48} & \underline{0.47} & \underline{0.54} & \underline{0.39} & \underline{0.29} & \underline{0.18} & \textbf{1.23} & \textbf{1.32} \\
ConfGF & 0.55 & 0.55 & 0.45 & 0.31 & 0.22 & 0.12 & \textit{div.} & \underline{1.54} \\
\bottomrule
\end{tabular}
\caption{Benchmark on angular error (RMSE$_{\mathrm{ang.}}$, rad) and all-atom positional error (RMSD$_{\mathrm{pos.}}$, \AA). The four middle columns report the fraction of samples whose angular error falls below different thresholds $\tau$. div. = diverged, with median computed excluding diverged samples. \textbf{Best} values are in bold and \underline{second best} are underlined. Estimates of the standard errors are given in Tab. \ref{tab:si_error_estimation}}
\label{tab:angular_and_rmse_benchmark}
\end{table}

To demonstrate the performance of \name{}, we show that (i) the model produces more accurate coordination environments compared to cheminformatics methods and other generative models. We further (ii) show that \name{} generates transition metal complex geometries with quantum-mechanical properties similar to those of the reference ground truth structure and (iii) exemplify \name{}’s performance on representative systems relevant to catalysis, drug design and functional materials.

We first quantify the accuracy of the generated coordination environments, \textit{i.e.}, the geometric arrangements of ligands around the metal center, by computing the root-mean-square error of the relative angles of ligating atoms (RMSE$_{\mathrm{ang.}}$) between the generated structure and the ground-truth structure (see Sec. \ref{si:rmse_ang} for definition). This metric specifically captures coordination geometry fidelity of interest for transition metal complexes, in contrast to all-atom RMSD, which reflects global structural similarity and is more influenced by changes distant to the metal center. We benchmark \name{} against RDKit's ETKDG, a widely used cheminformatics algorithm, and generative models from the literature; ConfGF and GeoDiff. 

As shown in Table \ref{tab:angular_and_rmse_benchmark}, \name{} achieves the lowest angular errors across all methods, demonstrating its capability to generate accurate coordination environments and ligand placements.
Specifically, \name{} shows a median angular error (RMSE$_{\mathrm{ang.}}$) of 0.41~rad, outperforming RDKit (0.66~rad) and ConfGF (0.55~rad), and slightly improving over GeoDiff (0.47~rad). Beyond average errors, \name{} consistently yields a substantially higher fraction of high-quality coordination environments. In particular, 41\% of generated structures lie below 0.3~rad ($\sim$17°), compared to 29\% for GeoDiff and 10\% for RDKit. At the more stringent threshold of 0.2~rad ($\sim$11.5°), \name{} still generates 31\% of samples below the cutoff, compared to 18\% for GeoDiff and 4\% for RDKit.
To assess the robustness of these estimates, we computed standard errors as shown in  Table~\ref{tab:si_error_estimation}. The resulting uncertainties are small relative to the observed performance differences ($\sim0.01$--$0.02$), indicating that the improvements in coordination geometry generation are robust. 
The low performance of RDKit can be attributed to the default coordination environment and random placement of ligands on available sites when using the SMILES extracted with xyz2mol without any prior stereogenic information. Without the RDKit initialization,  the performance of \name{} on the angular coordination environment metrics is reduced, but remains on par with GeoDiff, with, \textit{e.g.}, a median angular error 0.47 rad, equal to GeoDiff (see Tab. \ref{si:tab:ours_random_init_angular}).
In terms of heavy-atom RMSD, GeoDiff achieves the best performance (mean 1.23~Å), while \name{} attains a higher RMSD (2.3~Å) comparable to RDKit (2.3~Å, see Table \ref{tab:angular_and_rmse_benchmark}). 
The comparably lower performance of \name{} on heavy-atom RMSD can be attributed to the challenging learning of the rotational degrees of freedom, \textit{i.e.}, SO(3) diffusion. 
This can be seen in the training dynamics shown in SI Fig.~\ref{fig:loss-comparison}, where the normalised rotational loss \cite{leach_denoising_2022} remains significantly higher ($\ell_{\mathrm{rot}} \approx 0.85$) than the sphere-manifold loss introduced in this work ($\ell_{\mathrm{sphere}} \approx 0.25$).
Errors in learning the SO(3) \cite{leach_denoising_2022} rotational degrees of freedom, which show the highest loss in training and evaluation, lead to larger contributions to RMSD away from the metal center due to the rotation axis going through the coordinating atom (see Sec.  \ref{sec:methods_coupling_with_DOFs}). 
However, for relevant properties of transition metal complexes, we argue that the coordination environment often plays a dominant role in determining quantum mechanical properties, which we show in the following paragraph. 

\begin{table}[t]
\centering
\small
\setlength{\tabcolsep}{4pt}
\begin{tabular}{lccccc}
\toprule
Model & $|\mu|\,(\downarrow)$ & $\Delta E_{\mathrm{HOMO\text{-}LUMO}}\,(\downarrow)$ & $|\mu_{\mathrm{TM}}|\,(\downarrow)$ & $q_{\mathrm{TM}}\,(\downarrow)$ & $\mathrm{Tr}(Q_{\mathrm{TM}})\,(\downarrow)$ \\
 & (D) & (eV) & (D) & (a.u.) & (a.u.) \\
\midrule
\textbf{\name{} (ours)}& \textbf{1.73} & \textbf{1.06} & \uline{0.32} & \uline{0.42} & \textbf{0.66} \\
RDKit & \uline{2.11} & 1.27 & 0.50 & 0.66 & 0.91 \\
GeoDiff & 4.85 & \uline{1.18} & \textbf{0.31} & \textbf{0.38} & \uline{0.69} \\
ConfGF & 8.46 & 1.33 & 0.52 & 0.74 & 0.77 \\

\bottomrule
\end{tabular}
\caption{Mean absolute errors of electronic and transition-metal properties on the tmQMg benchmark, unrelaxed. Lower values are better ($\downarrow$). Best values are highlighted in bold, second-best underlined.}
\label{tab:properties_unrelaxed}
\end{table}

\begin{table}[b]
\centering
\small
\setlength{\tabcolsep}{4pt}
\begin{tabular}{lccccc}
\toprule
Model & $|\mu|\,(\downarrow)$ & $\Delta E_{\mathrm{HOMO\text{-}LUMO}}\,(\downarrow)$ & $|\mu_{\mathrm{TM}}|\,(\downarrow)$ & $q_{\mathrm{TM}}\,(\downarrow)$ & $\mathrm{Tr}(Q_{\mathrm{TM}})\,(\downarrow)$ \\
 & (D) & (eV) & (D) & (a.u.) & (a.u.) \\
\midrule
\textbf{\name{} (ours)}& \textbf{1.99} & 1.18 & 0.43 & 0.42 & \uline{0.84} \\
RDKit & \uline{2.21} & \textbf{0.78} & \uline{0.39} & \textbf{0.31} & 0.90 \\
GeoDiff & 2.70 & \uline{1.17} & \textbf{0.36} & \uline{0.35} & \textbf{0.77} \\
ConfGF & 8.49 & 1.46 & 0.76 & 0.60 & 1.23 \\

\bottomrule
\end{tabular}
\caption{Mean absolute errors of electronic and transition-metal properties on the tmQMg dataset, relaxed. Lower values are better ($\downarrow$). Best values are highlighted in bold, second-best underlined.}
\label{tab:properties_relaxed}
\end{table}

We now evaluate the generated structures based on quantum-mechanical properties computed with GFN2-xTB.
We show that \name{} generates transition metal complex geometries with quantum-mechanical properties close
to the reference ground-truth structures, as evaluated by GFN2-xTB calculations on generated and relaxed samples (see Tabs. \ref{tab:properties_unrelaxed} and \ref{tab:properties_relaxed}).
As properties, we consider both molecular properties (HOMO--LUMO gap, dipole moment), and atomic properties of the transition metal center, such as the atomic charge $q_{\mathrm{TM}}$, the magnitude of the atomic dipole $\left|\vec{\mu}_{\mathrm{TM}}\right|$ and quadrupole Tr$(Q_{\mathrm{TM}})$ moments.
While the HOMO--LUMO gap serves as a proxy for molecular reactivity and influences optical properties in materials, the metal charge, dipole and quadrupole moments characterize the local electron distribution.
As shown in Table~\ref{tab:properties_unrelaxed}, \name{} achieves the lowest or second-lowest errors across all properties. In particular, it attains the best performance for the dipole moment (1.73~D) and the HOMO--LUMO gap (1.06~eV), improving over RDKit (2.11~D, 1.27~eV) and GeoDiff (4.85~D, 1.18~eV). 
For metal-centered quantities, \name{} closely matches the best-performing model. After relaxation with GFN2-xTB, the comparable performance of \name{} is reduced, but remains within the range of the diffusion baselines (see Table~\ref{tab:properties_relaxed}).
Importantly, \name{} requires only 20 model evaluations at inference time, compared to 5000 steps for GeoDiff and ConfGF, making it substantially more efficient and enabling efficient large-scale geometry generation.

\begin{figure}[t!]
    \centering
    \includegraphics[width=\textwidth]{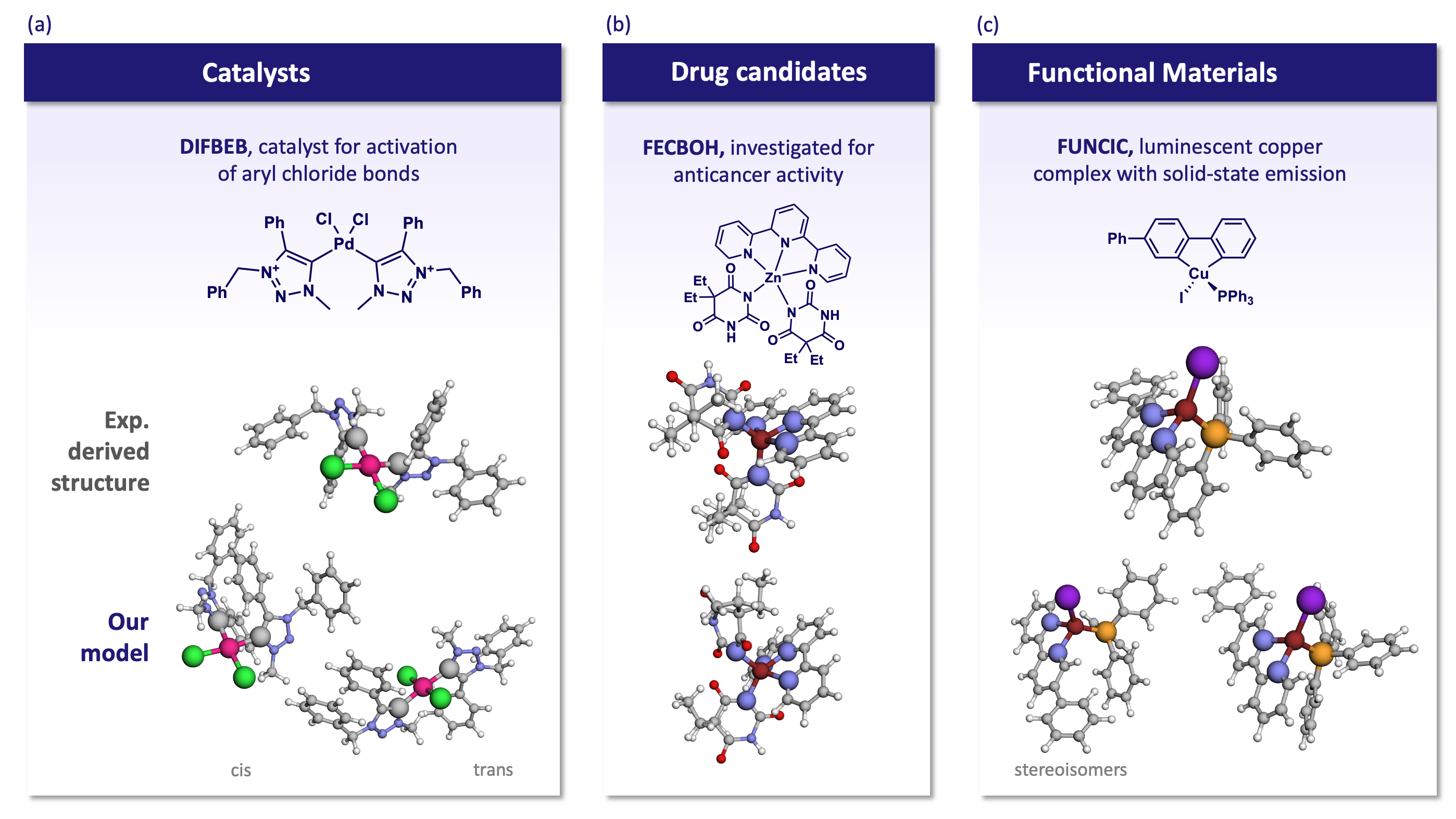}

    \caption{Representative examples of generated transition metal complexes with applications in (a) catalysis, (b) drug discovery and (c) functional materials.}
    \label{fig:results_selected_examples}
\end{figure}

Fig. \ref{fig:results_selected_examples} highlights test-set complexes relevant to (a) catalysis, (b) drug design, and (c) photochemistry, selected using application-specific keywords extracted via natural language processing \cite{kevlishvili_leveraging_2025}.
These examples, named in the following after their CSD entry, illustrate the ability of \name{} to sample stereoisomers --- including enantiomers and \textit{cis}-\textit{trans} isomers --- and to handle complex multidentate ligands. 
DIFBEB (Fig. \ref{fig:results_selected_examples}a) is a Pd complex relevant to homogeneous catalysis \cite{hohloch_abnormal_2013} and is reported to possess both \textit{cis} and \textit{trans} geometry. \name{} samples both the \textit{cis} and \textit{trans} isomers. This demonstrates the model’s ability to recover functionally relevant isomeric ensembles rather than a single minimum-energy structure.
FECBOH (Fig. \ref{fig:results_selected_examples}b) exemplifies complexes investigated for anticancer activity, motivated by the clinical success of  cisplatin along with oxaliplatin and carboplatin as anticancer drugs, which has raised significant interest in developing transition metal complexes with DNA/protein binding ability \cite{yilmaz_synthesis_2017}. This example illustrates the ability of \name{} to generate geometries containing multiple  bulky ligands, including a tridentate one, around a five-coordinate metal center.
FUNCIC (Fig. \ref{fig:results_selected_examples}c) represents Cu(I) complexes of interest in photochemistry due to their favorable photophysical properties, including solid-state emission and ligand-centered excited states. These features make such compounds attractive for applications in luminescent sensors, optoelectronic devices, and solar energy conversion, underscoring \name{}’s applicability to electronically and optically active transition-metal systems \cite{safin_luminescent_2015}. In this tetrahedral complex, \name{} generates both enantiomers (see Fig. \ref{fig:SI-FUNCIC-enantiomers} for visualization), demonstrating its ability to capture stereochemical diversity in chiral coordination environments.

\section{Conclusion}
We introduced \name{}, a manifold diffusion model designed to efficiently and accurately generate three-dimensional geometries of transition metal complexes (TMCs). By restricting the generative process to chemically meaningful internal coordinates --- specifically, the angular metal coordination environment and ligand torsions and rotations --- \name{} achieves high geometric and quantum-chemical fidelity while requiring few inference steps. This makes it a scalable and data-efficient alternative to Euclidean generative models, particularly well-suited to the limited-data regime characteristic of TMC datasets. Our approach can be extended for conditional generation based on target properties, enabling control over stereochemistry (\textit{e.g.}, \textit{cis}/\textit{trans} or axial/equatorial configurations) to optimize desired electronic or catalytic properties. This could be achieved for example by classifier-free guidance; or other model adaptation and finetuning methods \cite{domingo-enrich_adjoint_2024,gutjahr_constrained_2025, jensen_value_2026}. Here, \name{ } would guarantee valid metal-ligand bond lengths even under domain shift incurred under finetuning. Therefore, \name{} opens up the possibility for the efficient inverse design of transition metal complexes in a variety of applications across catalysis, drug design and materials science.

\section*{Data availability}
We used the tmQMg dataset from Kneiding et al. \cite{kneiding_deep_2023}. Our processed version is available at Zenodo (\href{https://doi.org/10.5281/zenodo.20072262}{10.5281/zenodo.20072262}).

\section*{Code availability}
The code used to train and evaluate the model is available at GitHub (\href{https://github.com/digital-chemistry-laboratory/TMCgen}{https://github.com/digital-chemistry-laboratory/TMCgen}) and Zenodo (\href{https://doi.org/10.5281/zenodo.20072262}{10.5281/zenodo.20072262}).

\section*{Competing interests}
The authors declare no competing interests.

\section*{Author contributions}
L.S. developed the methodology, implemented the model, and performed the experiments and data analysis. 
K.J. supervised the project and contributed to the conceptual design. 
L.S. and K.J. jointly wrote and revised the manuscript.

\section*{Acknowledgements}
We thank Vignesh Ram Somnath and Andreas Krause for constructive discussions. This publication was created as part of NCCR Catalysis (grant numbers 180544 and 225147), a National Centre of Competence in Research funded by the Swiss National Science Foundation.

\bibliography{references}
\newpage

\appendix
\section*{Supplementary Information}

\setcounter{figure}{0}
\setcounter{table}{0}
\renewcommand{\thefigure}{SI \arabic{figure}}
\renewcommand{\thetable}{SI \arabic{table}}


\section{Manifold diffusion: extended formulation}

\subsection{Generalization of the heat kernel to the \texorpdfstring{$n$}{n}-sphere}\label{si:nsphere}
The heat kernel on $S^n$ (Eq.~\ref{eq:methods:heat_kernel} in the main text for $n=2$) generalizes to arbitrary $n \geq 2$ by replacing the Legendre polynomials with Gegenbauer (ultraspherical) polynomials $C_\ell^{(\alpha)}$ with $\alpha = (n-1)/2$~\cite{zhao_exact_2018}:
\begin{equation}
G_n(\theta, \sigma(t)) = \frac{1}{\omega_n} \sum_{\ell=0}^{\infty} e^{-\ell(\ell+n-1)\sigma(t)^2 / 2}\,\frac{2\ell + n - 1}{n-1}\, C_\ell^{(\alpha)}(\cos\theta),
\end{equation}
where $\omega_n = 2\pi^{(n+1)/2}/\Gamma((n+1)/2)$ is the surface area of $S^n$. For $n=2$, $C_\ell^{(1/2)} = P_\ell$ recovers Eq.~\ref{eq:methods:heat_kernel}. The corresponding conditional density is obtained by weighting $G_n$ with the surface area element $\sin^{n-1}(\theta)$.
Gegenbauer polynomials are evaluated through the recurrence
\begin{equation}
\ell\, C_\ell^{(\alpha)}(x) = 2(\ell+\alpha-1)\,x\,C_{\ell-1}^{(\alpha)}(x) - (\ell+2\alpha-2)\,C_{\ell-2}^{(\alpha)}(x),
\end{equation}
with $C_0^{(\alpha)} = 1$ and $C_1^{(\alpha)}(x) = 2\alpha x$.

\section{Model architecture and training}

\subsection{Hyperparameters}\label{si:hyperparams}

Tables~\ref{tab:hparams-arch}--\ref{tab:hparams-sampling} list the hyperparameters used to train and sample from the E3NN score model for sphere--rotation--torsion diffusion on tmQMg. 

\begin{table}[h]
\centering
\begin{tabular}{ll}
\toprule
\textbf{Hyperparameter} & \textbf{Value} \\
\midrule
Number of convolutional layers & 4 \\
Scalar features ($n_s$) & 48 \\
Vector features ($n_v$) & 10 \\
Spherical harmonics ($\ell_{\max}$) & 2 \\
Activation & ReLU \\
Dropout & 0.1 \\
\bottomrule
\end{tabular}
\caption{Architecture of the E3NN score model.}
\label{tab:hparams-arch}
\end{table}

\begin{table}
\centering
\small
\begin{tabular}{ll}
\toprule
\textbf{Hyperparameter} & \textbf{Value} \\
\midrule
Distance embedding dim. & 64 \\
Center distance embedding dim. & 64 \\
Cross distance embedding dim. & 64 \\
Sigma embedding dim. & 32 \\
Time embedding type & sinusoidal \\
Time embedding scale & 10{,}000 \\
Position embedding type & sinusoidal \\
Position embedding scale & 10{,}000 \\
\bottomrule
\end{tabular}
\caption{Embedding dimensions and types.}
\label{tab:hparams-embed}
\end{table}

\begin{table}
\centering
\small
\begin{tabular}{ll}
\toprule
\textbf{Hyperparameter} & \textbf{Value} \\
\midrule
Cutoff (lower / upper) & 0 / 5 \\
Max neighbours & 30 \\
Center max radius & 30 \\
\bottomrule
\end{tabular}
\caption{Graph construction parameters.}
\label{tab:hparams-graph}
\end{table}

\begin{table}
\centering
\small
\begin{tabular}{ll}
\toprule
\textbf{Hyperparameter} & \textbf{Value} \\
\midrule
Rotation $\sigma_{\min},\sigma_{\max}$ & 0.02, 1.65 \\
Torsion $\sigma_{\min},\sigma_{\max}$ & 0.0314, 3.14 \\
Sphere $\sigma_{\min},\sigma_{\max}$ & 0.08, 1.55 \\
\bottomrule
\end{tabular}
\caption{Diffusion noise schedule.}
\label{tab:hparams-diff}
\end{table}

\begin{table}
\centering
\small
\begin{tabular}{ll}
\toprule
\textbf{Hyperparameter} & \textbf{Value} \\
\midrule
$w_{\text{rot}}$ & 1 \\
$w_{\text{tor}}$ & 1 \\
$w_{\text{sphere}}$ & 2 \\
\bottomrule
\end{tabular}
\caption{Loss weights.}
\label{tab:hparams-loss}
\end{table}

\begin{table}
\centering
\small
\begin{tabular}{ll}
\toprule
\textbf{Hyperparameter} & \textbf{Value} \\
\midrule
Optimiser & Adam \\
Learning rate & $1\times10^{-4}$ \\
Weight decay & 0 \\
Gradient clip & 0.8 \\
EMA decay & 0.999 \\
\bottomrule
\end{tabular}
\caption{Optimisation settings.}
\label{tab:hparams-opt}
\end{table}

\begin{table}
\centering
\small
\begin{tabular}{ll}
\toprule
\textbf{Hyperparameter} & \textbf{Value} \\
\midrule
Inference steps & 20 \\
\bottomrule
\end{tabular}
\caption{Sampling / inference settings.}
\label{tab:hparams-sampling}
\end{table}

\section{Metric definitions}

\subsection{Angular RMSE}\label{si:rmse_ang}

To assess the fidelity of the generated coordination environment, we evaluate the root-mean-square error of the pairwise angles formed by the ligating atoms at the transition metal (TM) center. Let $\mathbf{r}_i = \mathbf{x}_i - \mathbf{x}_{\mathrm{TM}}$ denote the position of ligating atom $i$ relative to the metal center, and let
\begin{equation}
    \theta_{ij} = \arccos\!\left( \frac{\mathbf{r}_i \cdot \mathbf{r}_j}{\lVert \mathbf{r}_i \rVert\, \lVert \mathbf{r}_j \rVert} \right)
\end{equation}
be the angle subtended at the TM by the pair $(i,j)$. The angular RMSE between a generated structure and its ground-truth reference is then defined as
\begin{equation}
    \mathrm{RMSE}_{\mathrm{ang.}} = \sqrt{ \frac{1}{N_{\mathrm{pairs}}} \sum_{i<j} \left( \theta_{ij}^{\mathrm{pred}} - \theta_{ij}^{\mathrm{true}} \right)^{2} },
    \qquad N_{\mathrm{pairs}} = \binom{N}{2},
\end{equation}
where the sum runs over all unordered pairs of the $N$ ligating atoms. By construction, this metric is invariant to global translations and rotations of the complex.

\subsection{Multidentate ligands}
Multidentate ligands have bite angles, which according to the terminology introduced by \cite{jing_torsional_2022} count towards the local environment and are kept fixed during training and inference, for example using reference angles based of RDKit or libraries such as tmQMg-L  \cite{kneiding_directional_2024}.
Following this approach, in the results shown in the main text, we keep the bite angles fixed with values from RDKit. However, the publicly available implementation allows choosing the bite angles to be diffused over, \textit{i.e.}, treating them like other torsion angles. The ligating atom which is used for the sphere diffusion is chosen randomly in the case of multidentate ligands.

\section{Additional results}
\subsection{Comparison of loss magnitudes}\label{si:loss-comparison}

\begin{figure}[h!]
    \centering
    \includegraphics[width=0.45\textwidth]{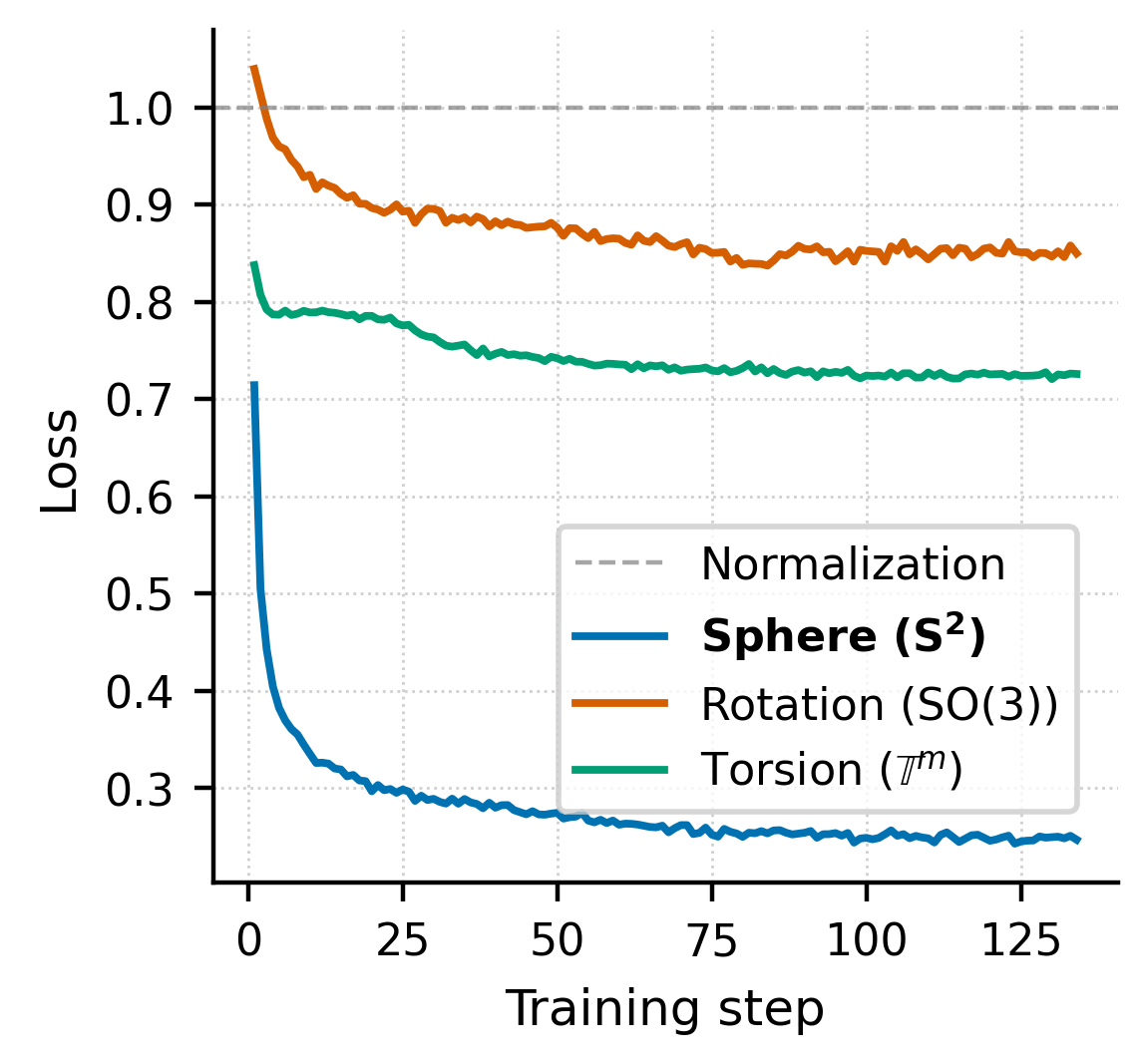}
    \caption{Per-component training losses (sphere, rotation, torsion) over the first 130 epochs, each normalised. The sphere loss introduced in this work converges quickly to ${\sim}0.25$, while the rotational loss \cite{leach_denoising_2022} plateaus near $0.85$.}
    \label{fig:loss-comparison}
\end{figure}

Fig. ~\ref{fig:loss-comparison} compares the per-component training losses (sphere, rotation, torsion), each normalised. The sphere loss  converges rapidly to roughly $0.25$ of its normalized values, whereas the rotational loss plateaus near $0.85$, indicating that rotational alignment is the dominant bottleneck and that the residual error in our generated structures is driven primarily by the rotational component.

\subsection{Additional results for different initialization scheme}\label{si:res_random_init}
With random initialization of the structures, especially random ligand rotations, \name{} remains stable but shows a moderate performance drop compared to the performance presented in the main text (see Tabs. \ref{si:tab:ours_random_init_angular} and  \ref{si:ourmethod_random_init_properties}). The angular RMSE increases from 0.43 to 0.51 rad, and RMSD rises from 2.31 Å to 2.38 Å, indicating a slightly less accurate  generation. Electronic property errors also degrade mildly (\textit{e.g.}, HOMO--LUMO gap from 1.06 eV to 1.31 eV, unrelaxed), while overall trends across metrics remain consistent, suggesting that the model does not collapse under fully uninformative starting geometries.

This result indicates that \name{} can refine randomly initialized complexes into chemically reasonable structures, but benefits from a chemistry-informed initialization (\textit{e.g.}, RDKit ETKDG) for optimal performance.
\begin{table}[t]
\centering
\small
\setlength{\tabcolsep}{4pt}
\begin{tabular}{lcccccccc}
\toprule
& \multicolumn{2}{c}{RMSE$_{\mathrm{ang.}}$ (rad)} & \multicolumn{4}{c}{Frac. RMSE$_{\mathrm{ang.}}$ $<\tau$} & \multicolumn{2}{c}{RMSD$_{\mathrm{pos.}}$ (\AA)} \\
\cmidrule(lr){2-3} \cmidrule(lr){4-7} \cmidrule(lr){8-9}
Method & Mean(↓) & Med.(↓) & $<0.5$(↑) & $<0.4$(↑) & $<0.3$(↑) & $<0.2$(↑) & Mean(↓) & Med.(↓) \\
\midrule
\name{} (random init.) & 0.51 & 0.47 & 0.54 & 0.43 & 0.31 & 0.18 & 2.38 & 2.32 \\
\bottomrule
\end{tabular}
\caption{Results of \name{} with random initialization on angular error (RMSE$_{\mathrm{ang.}}$, rad) and all-atom positional error (RMSD$_{\mathrm{pos.}}$, \AA). The four middle columns report the fraction of samples whose angular error falls below different thresholds $\tau$.}
\label{si:tab:ours_random_init_angular}
\end{table}

\begin{table}[t]
\centering
\small
\setlength{\tabcolsep}{4pt}
\begin{tabular}{lccccc}
\toprule
Model & $|\mu|\,(\downarrow)$ & $\Delta E_{\mathrm{HOMO\text{-}LUMO}}\,(\downarrow)$ & $|\mu_{\mathrm{TM}}|\,(\downarrow)$ & $q_{\mathrm{TM}}\,(\downarrow)$ & $\mathrm{Tr}(Q_{\mathrm{TM}})\,(\downarrow)$ \\
 & (D) & (eV) & (D) & (a.u.) & (a.u.) \\
\midrule
\multicolumn{6}{l}{\textit{Unrelaxed}} \\
\name{} (random initialization) & 2.65 & 1.31 & 0.57 & 0.78 & 0.89 \\
\addlinespace
\multicolumn{6}{l}{\textit{Relaxed}} \\
\name{} (random initialization) & 2.64 & 0.82 & 0.39 & 0.41 & 0.77\\
\bottomrule
\end{tabular}
\caption{Results of \name{} with random initialization of electronic and transition-metal properties on the tmQMg benchmark. Lower values are better ($\downarrow$). Best values are highlighted in bold, second-best underlined.}
\label{si:ourmethod_random_init_properties}
\end{table}

\subsection{Error estimation}

To assess the statistical robustness of the reported benchmark metrics,
uncertainties were estimated using a cluster bootstrap \cite{cameron_bootstrap-based_2008} over generated transition metal complexes ($B=10{,}000$ resamples).  We generated three samples per test complex while keeping the train/validation/test split and the trained model fixed. These uncertainty estimates therefore quantify variability arising from stochastic generation, but do not capture uncertainty due to changes in the training data, model initialization, or retraining of the model. Transition metal complexes were sampled with replacement, while retaining all associated samples for each selected transition metal complex (total of 3 samples), thereby preserving within-molecule correlations. Metrics were recomputed for each bootstrap replicate, and the standard deviation across replicates was reported as the standard error (SE). 

As shown in Table~\ref{tab:si_error_estimation}, the uncertainties are small for both angular and positional metrics. For example, the mean angular RMSE of \name{} is $0.43 \pm 0.01$~rad and the median angular RMSE is $0.41 \pm 0.02$~rad, indicating that the observed improvements over competing methods are substantially larger than the estimated statistical uncertainty. Positional RMSD metrics also show low uncertainty, with standard errors below $0.06$~\AA.

\begin{table}[t]\centering\small\setlength{\tabcolsep}{4pt}
\begin{tabular}{lcccccccc}\toprule
& \multicolumn{6}{c}{RMSE$_{\mathrm{ang.}}$ (rad)} & \multicolumn{2}{c}{RMSD$_{\mathrm{pos.}}$ (\AA)} \\
Method & Mean(↓) & Med.(↓) & $<0.5$(↑) & $<0.4$(↑) & $<0.3$(↑) & $<0.2$(↑) & Mean(↓) & Med.(↓) \\ \midrule
\textbf{Our method} & $0.43_{\pm 0.01}$ & $0.41_{\pm 0.02}$ & $0.60_{\pm 0.02}$ & $0.49_{\pm 0.02}$ & $0.41_{\pm 0.02}$ & $0.31_{\pm 0.02}$ & $2.31_{\pm 0.04}$ & $2.24_{\pm 0.06}$ \\
\bottomrule\end{tabular}
\caption{Subscripts show $\pm 1$ SE from a cluster bootstrap over molecules ($B=10{,}000$ resamples).}
\label{tab:si_error_estimation}
\end{table}

\begin{figure}[h!]
    \centering
    \includegraphics[width=0.4\textwidth]{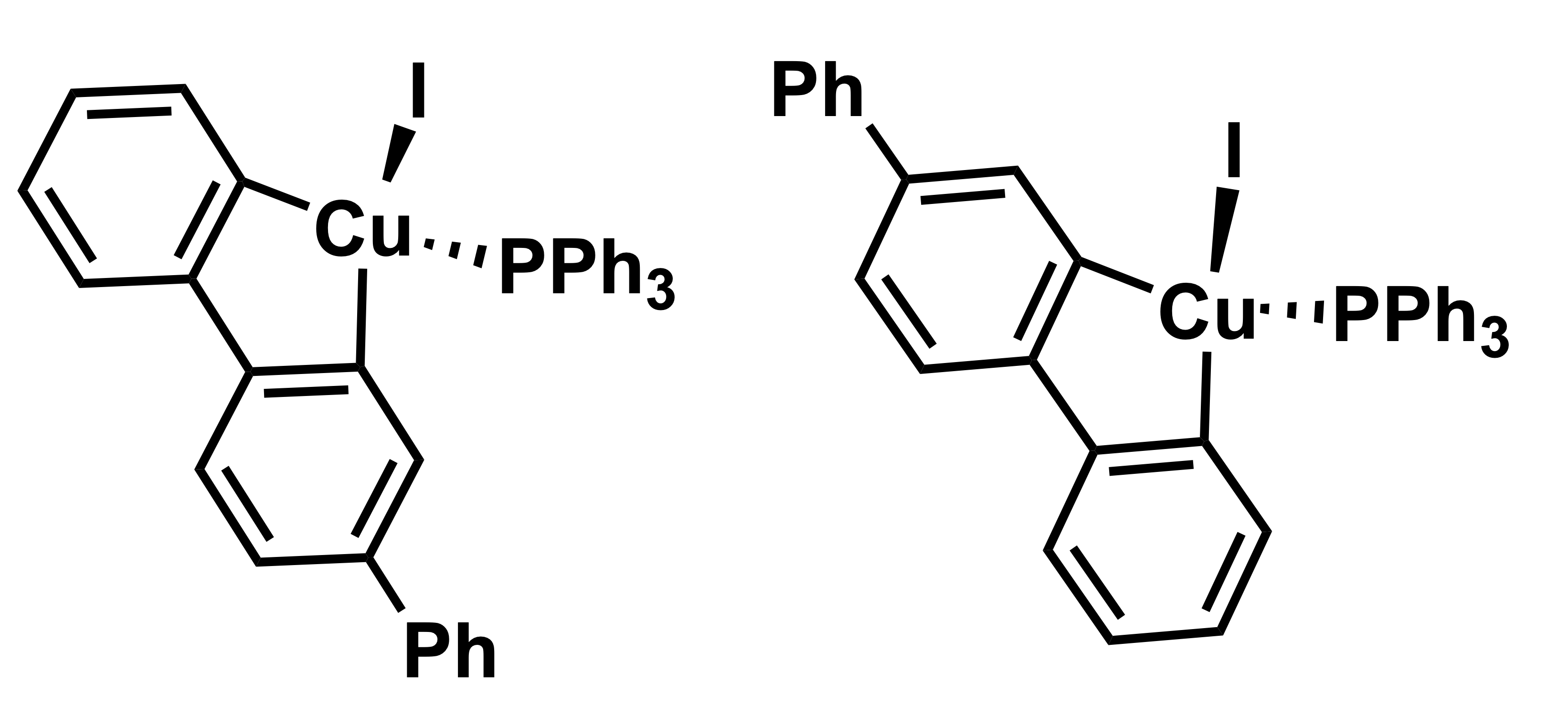}
    \caption{Enantiomers of FUNCIC}
    \label{fig:SI-FUNCIC-enantiomers}
\end{figure}

\end{document}